\documentclass[conference]{IEEEtran}
\IEEEoverridecommandlockouts

\usepackage{textcomp}
\usepackage{xcolor}
\usepackage{epsfig}
\usepackage{graphicx,amsfonts,amsmath,amssymb,amsxtra}
\usepackage{epstopdf}
\usepackage{comment}
\usepackage{rotating}
\usepackage{algorithmic}
\usepackage{graphicx}
\usepackage{tcolorbox}
\usepackage{multirow}
\usepackage[style=ieee, sorting=none]{biblatex}
\AtBeginBibliography{\small}

\def\BibTeX{{\rm B\kern-.05em{\sc i\kern-.025em b}\kern-.08em
    T\kern-.1667em\lower.7ex\hbox{E}\kern-.125emX}}

\begin{document}

\title{ \fontsize{24pt}{24pt}\selectfont
    Network Hexagons Under Attack: Secure Crowdsourcing of Georeferenced Data
    \thanks{
        This publication was developed as part of a program managed by Carnegie Mellon University Africa and supported by the MasterCard Foundation. The views expressed in this document are solely those of the authors and do not necessarily reflect those of the Carnegie Mellon University Africa or the MasterCard Foundation. This work was also partially funded by the Bill and Melinda Gates Foundation Carnegie through Upanzi Network at CMU-Africa.
    }   
}




\author{
    \IEEEauthorblockN{Okemawo Obadofin}
    \IEEEauthorblockA{
        \textit{Carnegie Mellon University, Africa} \\
        \textit{Carnegie Mellon University}\\
    }
    \and
    \IEEEauthorblockN{João Barros}
    \IEEEauthorblockA{
        \textit{Carnegie Mellon University, Africa} \\
        \textit{Carnegie Mellon University}\\
    }
}
\maketitle

\begin{abstract}
A critical requirement for modern-day Intelligent Transportation Systems (ITS) is the ability to collect geo-referenced data from connected vehicles and mobile devices in a safe, secure and anonymous way. The Nexagon protocol, which builds on the IETF Locator/ID Separation Protocol (LISP) and the Hierarchical Hexagonal Clustering (H3) geo-spatial indexing system, offers a promising framework for dynamic, privacy-preserving data aggregation. Seeking to address the critical security and privacy vulnerabilities that persist in its current specification, we apply the STRIDE and LINDDUN threat modelling frameworks and demonstrate, among other findings, that the Nexagon protocol is susceptible to user re-identification, session linkage, and sparse-region attacks. To address these challenges, we propose an enhanced security architecture that combines public key infrastructure (PKI) with ephemeral pseudonym certificates. Our solution guarantees user and device anonymity through randomized key rotation and adaptive geospatial resolution, thereby effectively mitigating re-identification and surveillance risks in sparse environments. A prototype implementation over a microservice-based overlay network validates the approach and underscores its readiness for real-world deployment. Our results show that it is possible to achieve the required level of security without increasing latency by more than 25\% or reducing the throughput by more than 7\%.

\end{abstract}

\begin{IEEEkeywords}
Secure Crowdsourcing, Mobile Networks, Privacy-Preserving Protocols, Threat Modeling for Geo-Privacy, Mobile Edge Security.
\end{IEEEkeywords}

\section{Introduction}
\begin{figure}[htbp]
\vspace{-10pt}
\centerline{\includegraphics[width=1.0\linewidth]{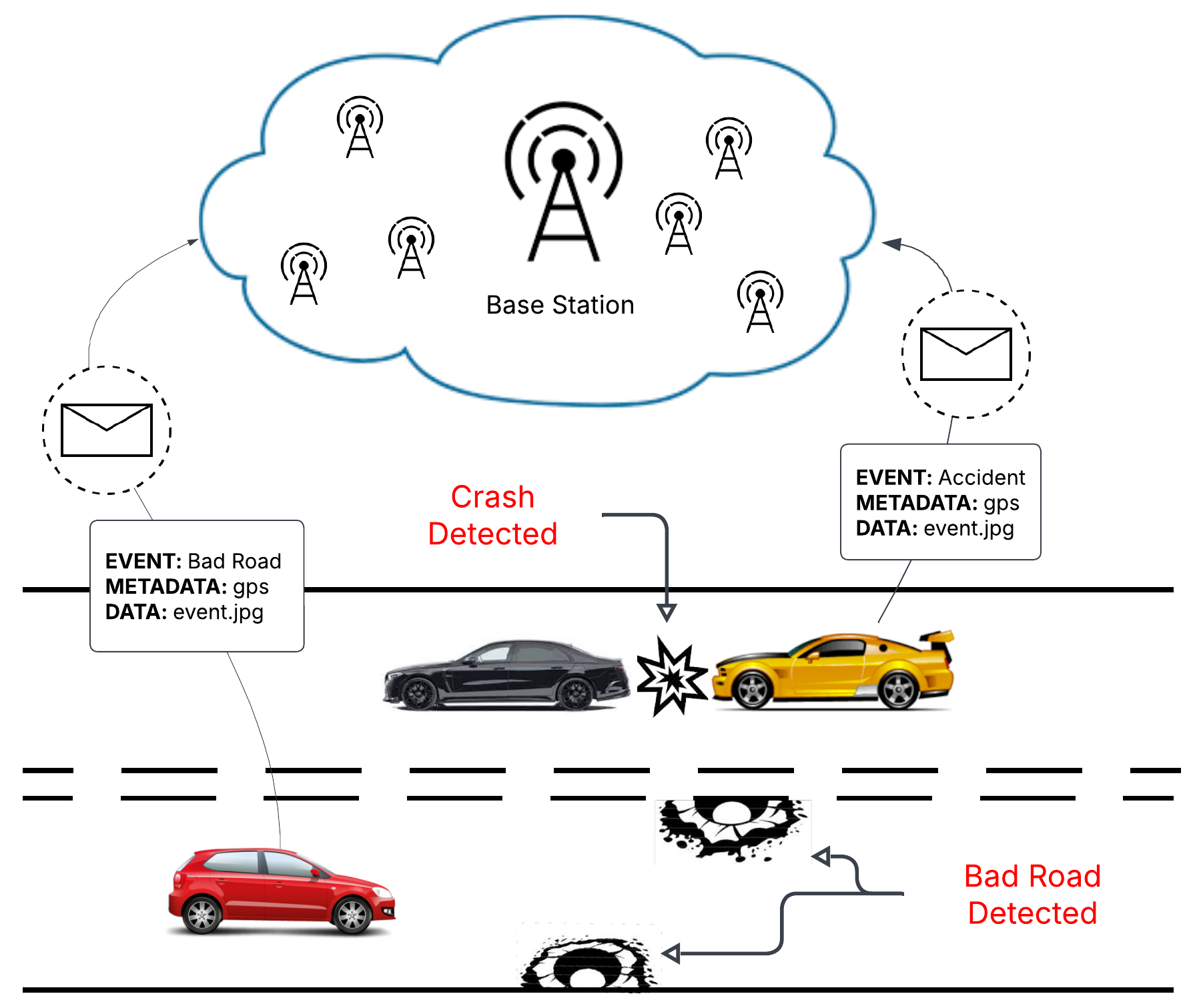}}
\caption{A traffic scene showing a network of connected vehicles capturing and reporting road conditions and vehicle crashes in real time.}
\label{Edge-devices-acquiring-data}
\vspace{-10pt}
\end{figure}

Distributed machine learning (ML) running on connected vehicles and mobile devices promises to deliver significant gains in transportation efficiency and sustainability, most notably by improving route planning and traffic optimization in real time ~\cite{1} with multiple layers of contextual information~\cite{2}.
As illustrated in Figure~\ref{Edge-devices-acquiring-data}, connected vehicles are now serving as edge devices capable of acquiring massive amounts of geo-referenced data for a variety of use cases, from road defects to crash detection.
The accuracy and granularity of such data ultimately determines the overall safety and performance of intelligent transportation systems \cite{4}. 

A key challenge in this class of data-driven solutions is how to collect real-time data from millions of connected vehicles and mobile edge devices, while addressing the very significant security and privacy concerns of all those who own and operate them in real-world situations. In other words, to ensure that no attacker is able to track or compromise any of the mobile nodes or their data, the underlying intelligent transportation ecosystem requires secure data aggregation systems that are implemented at scale~\cite{5, 6}. 

In light of these challenges, the Internet Engineering Task Force (IETF) is currently discussing and standardizing the use of network hexagons ("nexagons")~\cite{3} for secure mobile data collection. This so called \textit{Nexagon Protocol} builds on the Locator/ID Separation Protocol (LISP)~\cite{7158286} and enables a network of mobile nodes to map traffic signs, vehicle routes, construction works, natural hazards and other road conditions in real time. The protocol relies on geographically distributed agents to transmit sensor-derived attributes, while ensuring timely updates across the entire network. 


The Nexagon protocol lacks detailed guidelines for practical authentication mechanisms necessary to support its security requirements. Therefore, our research focuses on critically examining the protocol's architecture to identify aspects that undermine the security and privacy of connected vehicles and their users. Our goal is to conduct a systematic threat analysis and implement a real-world prototype with effective mitigations for identified vulnerabilities that informs future studies.



Our research addresses critical gaps in the Nexagon protocol by providing: (1) a comprehensive analysis of security and privacy threats within the Nexagon protocol, along with corresponding mitigations; (2) a PKI-based authentication system employing pseudonym certificates and secure key rotation to guarantee secure discovery and enhanced anonymity; and (3) a prototype Nexagon implementation along with performance metrics that demonstrates the feasibility and overhead of PKI-based authentication enhancements.

The remainder of the paper is structured as follows: Section II offers an overview of related research on privacy-preserving protocols within ITS context, leading into our contributions. Section III elaborates on the Nexagon architecture, detailing its association with the LISP and the Hierarchical Hexagonal Grid System (H3). Section IV conducts a security and threat analysis of the Nexagon protocol, identifying potential vulnerabilities, and proposing mitigations. Section V details our implementation strategy for the Nexagon protocol, followed by a comprehensive evaluation of our security attachments. Finally, Section VI concludes with a summary of our findings and suggests areas for future research.

\section{Related Work}
The heavy dependence of ITS applications on geo-referenced data introduces inherent privacy risks~\cite{10.1007/978-3-030-03405-4_5} for participating users. Providing user anonymity in such systems is essential to maintain public trust and encourage widespread adoption.
Research on k-anonymity~\cite{doi:10.1142/S0218488502001648} introduced a mathematically robust solution to address privacy risks by obscuring user identities. The anonymity model ensures that each individual's data is indistinguishable from at least k-1 other individuals in the dataset. This is achieved by introducing dummy users that enhance anonymity and provide statistical guarantees against re-identification. 
In contrast, differential privacy (DP)~\cite{10.1007/11787006_1} adopts a different strategy by adding randomness to the data through the introduction of noise, ensuring that the output of queries reveals minimal information about any individual while maintaining overall dataset utility.

While privacy-preserving techniques like k-anonymity show promising results, significant challenges persist when applied to real-time, dynamic environments like ITS. For instance, ~\cite{10.1007/11787006_1} used k-anonymity to obscure the actual location for users of location based services (LBS) by introducing dummy requests from users with spoofed locations. This approach ensures user anonymity but leads to inefficiencies in environments with infrastructural resource constraints. In ITS, where timely updates on routes and traffic conditions are critical, this method results in significant bandwidth consumption without proportional benefits. Additionally, mobile clients using this scheme for spoofing face performance degradation, as duplicate requests are required to maintain anonymity.

In contrast, differential privacy introduces controlled noise into the data, ensuring that individual user data remain protected without requiring any redundancy. However, the introduction of noise can obscure critical details, resulting in less precise traffic predictions, route optimization, or fleet management insights~\cite{10.1145/3035918.3054779}.

Several other research efforts have explored real-world solutions aimed at preserving privacy in mobile crowdsourcing systems. The Methods outlined by ~\cite{8485464, 8678475, PEREZ2022100450, https://doi.org/10.1155/2018/8959635}, safeguard user data and maintain anonymity while still allowing for efficient data aggregation and task allocation. However, these solutions either fall short when applied to dynamic and real-time environments or are too technically challenging to implement and maintain.
Existing approaches to guarantee anonymity do not meet the operational needs of ITS, highlighting the need for a more efficient, scalable, and adaptable solution. The Nexagon protocol~\cite{3} addresses these challenges by introducing a solution designed for real-time data exchange with security extensions tailored to distributed connected vehicles and other kinds of mobile edge devices.

\section{Network Hexagons: An Overview}
The Nexagon protocol establishes a distributed network of mobile edge devices designed to support real-time streaming of geo-referenced data while ensuring user privacy. The protocol leverages the advanced addressing semantics of the Locator/ID Separation Protocol (LISP)~\cite{7158286} for efficient client management and employs the H3 hierarchical spatial indexing system to effectively localize clients. LISP addresses the challenges of the modern internet by separating the concerns of uniquely identifying a device and its routing context within the network. H3 was developed to improve the user experience in applications that rely on location-based services~\cite{sym11060731, h3docs}.

As a key component of the Nexagon network, LISP provides a foundational architecture for localizing clients within hexagons. By separating global and local scopes, LISP mirrors traditional network designs through its use of Routing Locators (RLOCs) and End-Point Identifiers (EIDs). RLOCs enable communication between LISP-enabled edge routers, offering wide-area network access, while EIDs facilitate interactions among clients within the Nexagon overlay. To ensure efficient data forwarding and seamless network discovery, edge routers maintain mapping caches that store the associations between EIDs and RLOCs.

\begin{figure}[ht!]
\vspace{-5pt}
\centering
\includegraphics[width=1.0\linewidth]{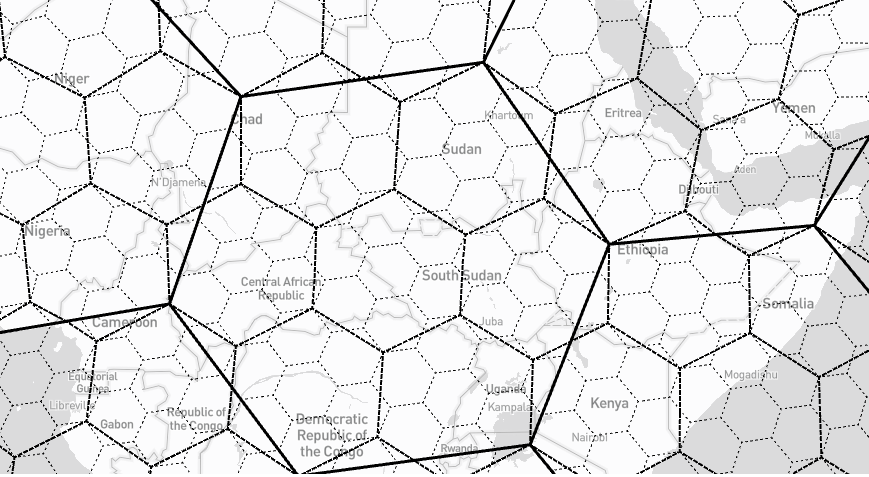}
\caption{Sectional map with hexagonal tiles of varying granularity superimposed showing how vehicles and mobile devices can be indexed.}
\label{nexagons}
\vspace{-5pt}
\end{figure}

The H3 package, developed by Uber Technologies, was designed to enhance location-based services for user applications. The library converts the GPS coordinates of clients into unique identifiers that correspond to a specific hexagon within a defined grid. Each hexagon, also known as a "H3 tile", groups devices with similar GPS coordinates in the same tile. The granularity of these tiles is determined by a parameter called resolution, which defines the level of detail for a given geographical area ~\cite{h3docs}. As illustrated in Figure~\ref{nexagons}, the sectional map visually demonstrates how H3 tiles establish boundaries, enabling vehicles to be grouped based on their geographic location.

\begin{figure}[ht!]
\vspace{-5pt}
\centering
\includegraphics[width=1.0\linewidth]{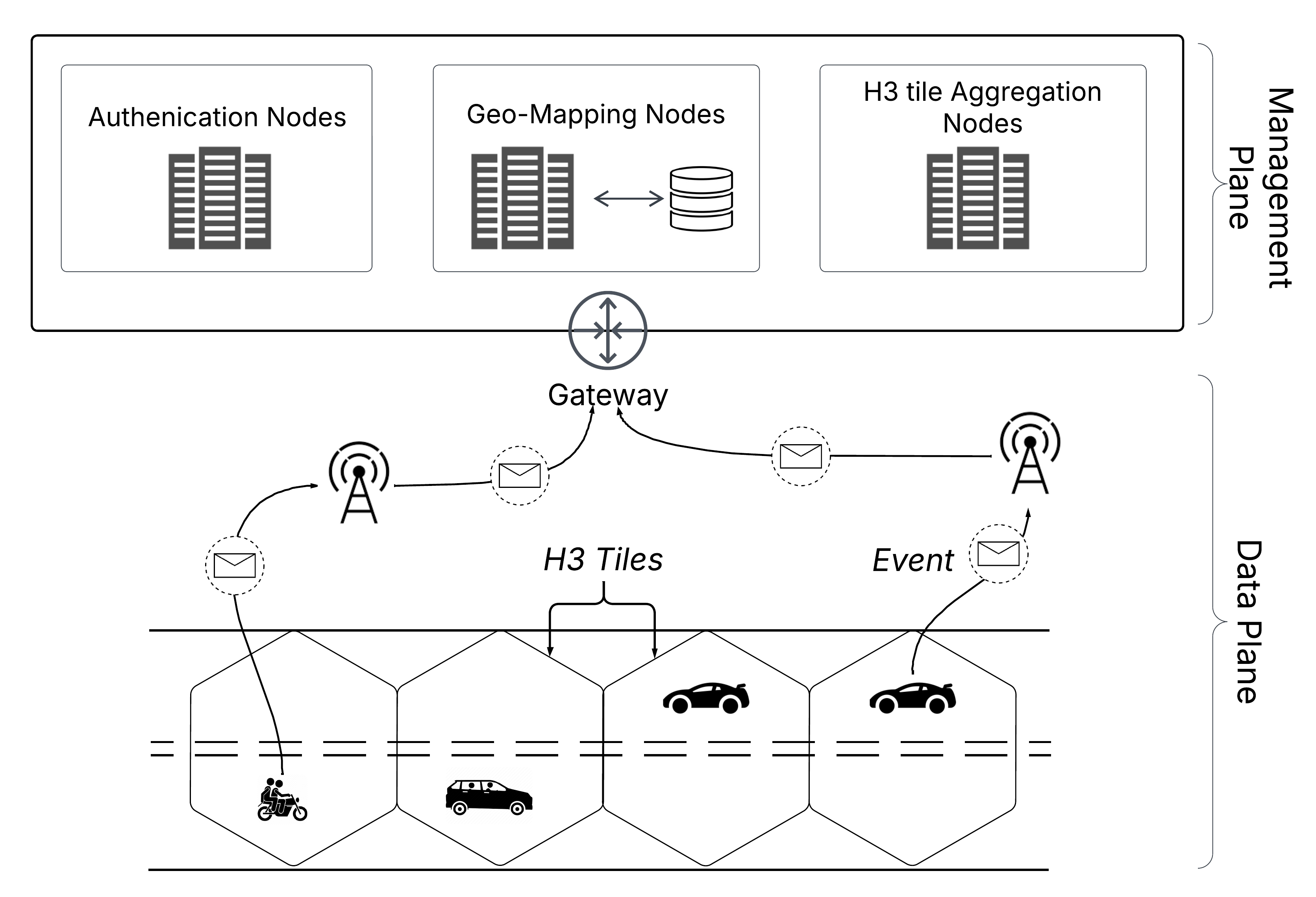}
\caption{An Overview of the Nexagon protocol architecture.}
\label{protocol-architecture}
\vspace{-5pt}
\end{figure}

In addition to mobile edge devices or clients, the Nexagon specification outlines several key components that enable and support the services it provides for edge devices. Figure~\ref{protocol-architecture} provides a comprehensive visual representation of all the participating elements within this scheme, highlighting their interactions with other components and their specific role in enabling secure, scalable, and privacy-preserving data aggregation. 

\begin{itemize}
    \item \textbf{Authentication Nodes:} Handles client association and onboarding into the Nexagon network by securely providing the necessary credentials through a specified method.
    \item \textbf{Geo-Mapping Nodes:} Maintains a spatial database that maps geo-referenced updates to corresponding client Endpoint Identifiers (EIDs).
    \item \textbf{H3 Aggregation Nodes:} Processes data collected within each H3 tile, enabling localized analysis and supporting a wide range of services for users. Aggregated results are stored in data lakes for future analytics and insights.
\end{itemize}

\section{Security and Privacy Analysis}
As it stands, the Nexagon protocol specification does not define the specific mechanism to be used for client authentication. To ensure the network is robustly protected against rogue actors, it is critical to conduct a comprehensive investigation of potential attack vectors and develop strong security mechanisms capable of mitigating these threats effectively. This analysis is aimed specifically for safeguarding the integrity and privacy of mobile edge devices. To achieve this, we apply the STRIDE~\cite{STRIDE} and LINDDUN~\cite{Deng2011APT} threat modeling frameworks. Together, these complementary techniques enable us to systematically identify privacy risks within the Nexagon protocol.

First, we decompose the Nexagon system into Data Flow Diagrams (DFDs), providing a visual representation of the interactions and information flows between its components. Next, we map threat categories to corresponding DFD elements, identifying components that are susceptible to security and privacy risks. Subsequently, attack tree templates from the STRIDE and LINDDUN frameworks are leveraged to uncover potential attack scenarios and develop mitigation strategies. Finally, we summarize our findings in a comprehensive table that outlines the identified threats, proposed mitigations, and privacy-enhancing solutions to be considered within our implementation setup.

\subsection{System Decomposition}
\begin{figure}[ht!]
    \centering
    \includegraphics[width=1.0\linewidth]{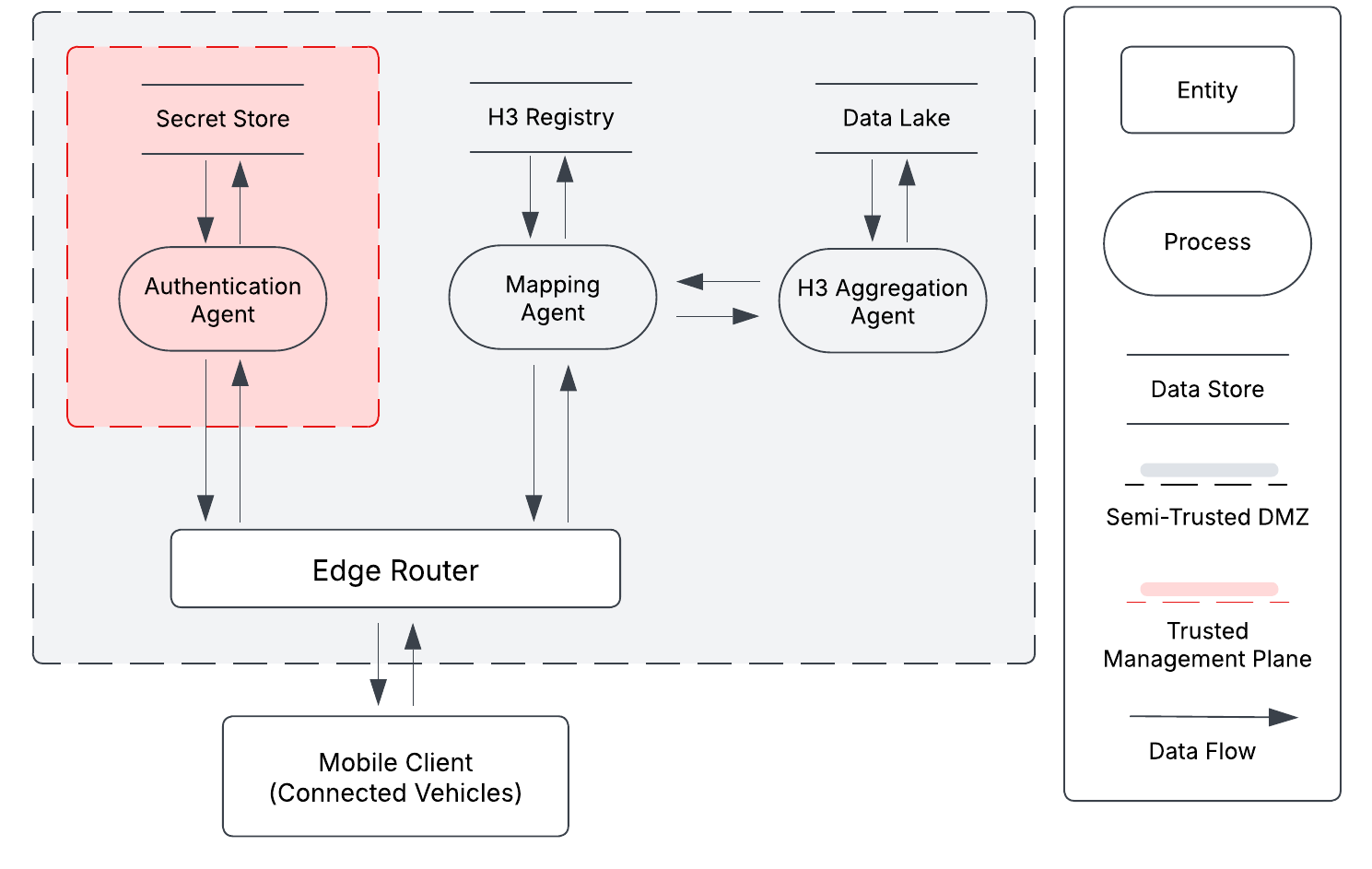}
    \caption{Data flow diagram for the Nexagon protocol.}
    \label{data flow diagram}
\end{figure}

To evaluate the security and privacy risks of the Nexagon protocol in a systematic way, we decompose its architecture into foundational elements using a DFD. Therefore, each element is represented as an entity, process, data store, or data flow. As shown in Figure \ref{data flow diagram}, the resulting trust boundaries are defined across three different zones: (1)~\textit{Untrusted External} (client processes and edge infrastructure), (2)~\textit{Semi-Trusted "Demilitarized Zone" (DMZ)} (aggregation processes and edge routers), and (3)~\textit{Trusted Management Plane} (core authentication processes). Whereas external entities like connected vehicles and other mobile edge devices are assigned to the Untrusted External Zone, edge routers are scoped to the Semi-Trusted DMZ, because they mediate the traffic between untrusted clients and trusted internal processes. The Authentication process operates within the Trusted Management Plane because it handles sensitive tasks such as issuing pseudonym credentials and managing cryptographic secrets. Data flows such as geo-referenced updates are scoped across trust boundaries.

\subsection{Mapping Threat Categories}
 Table \ref{dfd decomposition} provides a comprehensive mapping of security and privacy threat categories to their respective DFD elements in the Nexagon architecture. Each intersection signifies the applicability of a specific threat category to a particular DFD element:\textit{ entity, process, data store, or data flow}. This mapping highlights how different components of the system are exposed to varying risks and serves as a foundation for targeted mitigation strategies.

For instance, properties like \textit{Linkability} and \textit{Disclosure} span across multiple DFD elements, reflecting their pervasive nature in systems handling sensitive data. Similarly, threat categories like Spoofing and Tampering emphasize vulnerabilities in entities and processes, while the impact of Denial of Service is more prevalent to processes and data flows due to their role in maintaining system availability. Repudiation and Information Disclosure were excluded from the STRIDE mapping because they are already addressed within the LINDDUN framework. Similarly, Content Unawareness and Non-Compliance were excluded as they do not  impact the privacy of users in Nexagons.

\begin{table}[ht!]
    \caption{Mapped Threat Categories to DFD Elements}
    \centering
    \renewcommand{\arraystretch}{2.5} 
    \setlength{\tabcolsep}{4pt} 
    \begin{tabular}{|p{0.12\linewidth}|p{0.24\linewidth}|p{0.10\linewidth}|p{0.12\linewidth}|p{0.10\linewidth}|p{0.10\linewidth}|}
        
        \hline
        \multirow{2}{*}{} & \multirow{2.5}{*}{\shortstack[l]{\textbf{Threat} \textbf{Category}}} & \multicolumn{4}{|c|}{\textbf{DFD Element}}  \\

        \cline{3-6}
        & & Entity & Process & Data Store & Data Flow \\
        
        \hline
        \multirow{5}{*}{\textbf{\rotatebox{45}{LINDDUN}}} & Linkability & X & X & X & X \\    

        \cline{3-6}
        & Identifiability & X & X & X & X \\

        \cline{3-6}
        & Non-repudiation & & X & & X \\

        \cline{3-6}
        & Detectability & X & & & X \\

        \cline{3-6}
        & Disclosure & X & & X & X \\



        \hline
        \multirow{3}{*}{\textbf{\rotatebox{45}{STRIDE}}} & Spoofing & X & X & &  \\
        
        \cline{3-6}
        & Tampering & & X & X & \\

        \cline{3-6}
        & Denial of Service & & X & & X \\




        
        \hline
         
    \end{tabular}
    \label{dfd decomposition}
\end{table}

\subsection{Threats, Mitigations, and Privacy Enhancing Solutions}
In Table~\ref{threats and mitigations}, we present a comprehensive list of threats discovered during our systematic threat modeling process. For each identified threat, we used insights from attack trees provided by the LINDDUN framework to develop attack scenarios and propose mitigations based on evaluated privacy-enhancing solutions. This systematic approach ensures that both security vulnerabilities and privacy risks are addressed holistically. We summarize our findings by presenting two critical threat models that highlight vulnerabilities with high-risk factors. We address the vulnerabilities identified during our prototype implementation, as outlined in the following section. For brevity, we include a table summarizing attack scenarios, providing descriptions for other critical attacks not explicitly detailed.

\begin{table*}[ht!]
    \caption{Threats, Affected Components, Attack Descriptions, and Mitigations in Network Hexagons}
    \centering
    \renewcommand{\arraystretch}{1.6}
    \setlength{\tabcolsep}{4pt}
    \begin{tabular}{|p{0.12\linewidth}|p{0.15\linewidth}|p{0.28\linewidth}|p{0.28\linewidth}|p{0.10\linewidth}|}
        \hline
        \textbf{Threat} & \textbf{Affected \newline Component(s)} & \textbf{Attack Description} & \textbf{Mitigation} & \textbf{Risk Level} \\

        \hline
        Session Linkage & \textendash Mobile Client \newline \textendash Authentication Agent & An attacker links users by connecting credentials across different processes and tracking actions across sessions through static identifiers or similar patterns. & \textendash Use pseudonymized EIDs that are rotated dynamically~\cite{10.1145/586110.586137}. \newline \textendash Add dummy traffic to prevent timing-based correlation. & High \\

        \hline
        Request Profiling & \textendash Mobile Client & An attacker links user behavior across processes to identify patterns (for example request frequency). & \textendash Add noise to request patterns to obscure user behavior~\cite{chakravarty2014traffic}. \newline \textendash Route requests through mix networks to anonymize traffic. & Medium\\

        \hline
        Sparse Region \newline Attack & \textendash Mobile Client & An attacker exploits sparsely populated or remote areas by leveraging the low density of clients within static hexagonal grids. & \textendash Expand hexagonal regions dynamically in sparse areas. \newline \textendash Ensure at least k clients are indistinguishable in any region~\cite{chow2008spatial}. & High \\

        \hline
        User \newline Re-identification & \textendash Mobile Client \newline \textendash Mapping Agent & An attacker identifies users by correlating pseudonyms with external datasets (for example IP addresses). & \textendash Authenticate users without revealing identity. \newline \textendash Encrypt all metadata in communications and storage or use an onion router. & Medium \\
    
        \hline
        Forged Logs or \newline Audit Entries & \textendash Mobile Client \newline \textendash Authentication Agent \newline \textendash Mapping Agent \newline \textendash Aggregation Agent & A malicious actor modifies logs to deny responsibility for specific actions. & \textendash Use blockchain or append-only storage for audit trails. \newline \textendash Cryptographically sign logs to ensure integrity. & Low \\
  
        \hline
        Client Request/Response Replay & \textendash Authentication Agent \newline \textendash Mapping Agent & An attacker replays valid authentication requests to bypass non-repudiation mechanisms. & \textendash Include nonces in requests to prevent replay attacks. \newline \textendash Use mutual TLS (mTLS) for bidirectional verification. & Medium \\
        
        \hline
        User Data Leakage \newline and Eavesdropping &  \textendash Mobile Client \newline \textendash Authentication Agent & Attackers exploit weak encryption or poor access controls to intercept unencrypted transmissions (including geo-referenced data), exposing sensitive user information. & \textendash Encrypt all data in transit and at rest. \newline \textendash Enforce strict role-based access controls (RBAC). & Medium \\

        \hline
        Spoofed Agent &  \textendash Mobile Client \newline \textendash Mapping Agent  & An attacker impersonates a legitimate client or authentication agent. & \textendash Use hardware-backed attestation for device verification. \newline \textendash Require both parties to authenticate each other using certificates~\cite{9824568}. & High \\

        \hline

            \multicolumn{5}{|c|}{\multirow{3}{*}{\shortstack[l]{
                \textbf{High -} \hspace{20pt} Very strong likelihood of occurring and has a critical effect on the Nexagons, and \textbf{Not} currently addressed by well-known schemes \\ 
                \textbf{Medium -} \hspace{8pt} Very strong likelihood of occurring, has a critical effect on the Nexagons, and currently addressed by well-known schemes \\ 
                \textbf{Low-} \hspace{25pt} Very weak likelihood of occurring and has a critical effect on the Nexagons }}} \\

            \multicolumn{5}{|c|}{} \\ 

            \multicolumn{5}{|c|}{} \\ 

            \hline

    \end{tabular}
    \label{threats and mitigations}
    \vspace{-10pt}
\end{table*}

\subsubsection{Narrow-region attacks in hierarchical clustering}
\begin{figure}[ht!]
\centering
\includegraphics[width=0.8\linewidth]{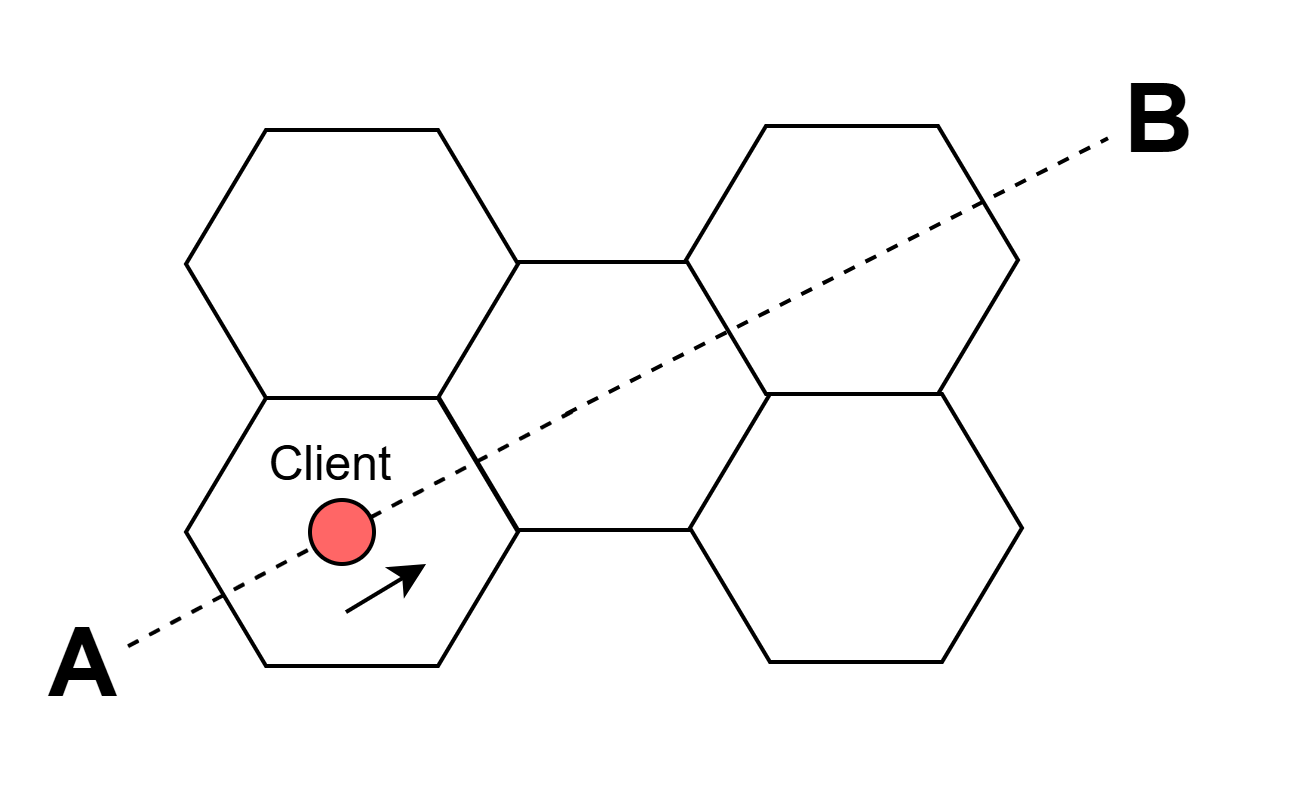}
\caption{A mock scenario illustrating a lone clients roaming within sparsely populated hexagons.}
\label{remote-client-risk}
\end{figure}

While hexagons offer an efficient approach to localizing data from mobile clients, privacy concerns arise when H3 tiles are overlaid on road networks in sparsely populated or remote areas. In such regions, the low density of clients can inadvertently expose individuals to privacy risks. As illustrated in Figure~\ref{remote-client-risk}, a lone client moves from Point \textbf{A} to Point \textbf{B} along a defined road with overlaid hexagonal tiles. As the client provides periodic updates, its position within the grid can be easily tracked. Each time the client enter a new hexagon, its presence can be easily noticed, making it a trivial task to infer the client's direction and ultimately predict the final destination.

One approach to mitigate this risk is to take advantage of the H3 modules that enable us to adjust how dense or coarse the localization of mobile edge devices should be. A parameter specified as the resolution ranging from 1 to 15 is set to control this feature. By dynamically altering the resolution in remote areas, we can achieve an effect similar to k anonymity, while maintaining the overall utility of geo-referenced data gathered. Varying the resolution in such scenarios to accommodate more clients will pose a difficult task for adversaries to decipher.

\vspace{5pt}
\subsubsection{Spoofed Control Plane Agent}
The Nexagons management agents serve as vital abstractions that oversee the network's overall state and manage core services handling packet delivery, data pre-processing, and authentication for mobile edge clients. Ensuring that clients communicate exclusively with legitimate agents is essential to maintain trust among entities. Malicious actors may attempt to disrupt the service discovery process or compromise client privacy during device initialization by impersonating a legitimate authentication server. To combat this threat, a trusted platform module (TPM)~\cite{5931361} can provide essential protection for onboarding mobile edge clients by securely generating, storing, and managing cryptographic keys and credentials. When a client attempts to authenticate in a nexagon, the TPM provides verifiable information, such as a server name to be resolved and secure cryptographic keys necessary to solve a challenge, ensuring that only legitimate clients can access management services. Previous research has demonstrated that software implementations of TPM, such as firmware-based TPM (fTPM) offer robust security guarantees comparable to dedicated TPM hardware and have been deployed in millions of mobile devices~\cite{raj2016ftpm}.

\section{Implementation and Performance Analysis}
The Nexagon~\cite{3} protocol specification does not explicitly define which layer of the internet stack the protocol should be implemented. However, since it leverages LISP, a Layer 3 protocol, it's development naturally aligns with the network layer. However, this approach requires significant changes to existing standards and configurations. Deploying the protocol as an overlay offers significantly more flexibility when integrating with existing infrastructure, promoting faster integrations and seamless compatibility with current systems~\cite{Muhammad2021, Fahmy2007}. Based on this observation, we now discuss in detail our approach for protocol's deployment as an overlay, outlining development processes for integrating mitigations for threats, and some architectural abstractions along with core technologies utilized.


\subsection{Deployment Architecture}
Leveraging the microservice architecture, each component was developed as a isolated services, enabling it to operate independently while interacting through specified APIs. Core functionalities were encapsulated within distinct containers, each representing a Nexagon agent. In summary, we complemented the Nexagon components with the following security enhancements:

\vspace{5pt}
\subsubsection{Authentication}
We designed the authentication process to operate as a root Certificate Authority (CA), serving as the single source of truth. However, an alternative deployment strategy could adopt a hierarchical CA structure with intermediate CAs to provide flexibility and scalability, particularly when accommodating existing vendors and cloud service providers. Intermediate CAs will act as delegated entities authorized by the root CA to issue certificates, which enhances security and operational efficiency~\cite{10671564}. While adhering to the core principles of PKI~\cite{7163785}, we augmented our solution with two essential mechanisms integrated into its operations.

\begin{itemize}
    \item Anonymity through randomness: To enhance security, we periodically rotate the keys used by the agent for signing, forcing all agents to re-authenticate periodically. This offers an additional layer of security compared to a traditional PKI deployments that maintain certificate validity for extended periods without requiring renewal or re-validation. A similar approach is applied to the EIDs generated by the client agent discussed in the following section. Additionally, a variable hexagonal resolution is employed after each key rotation for clients who are sampled by the management plane to be at risk of being located in a sparse region. This approach mitigates the privacy risks associated with the hierarchical clustering we previously highlighted.

    \item Pseudonym certificates for enhanced privacy: The use of pseudonym certificates mitigates the potential for unwanted surveillance and limits the exposure of sensitive data. This means that even if a certificate is compromised, its impact is minimized, as it cannot be linked back to the client. Pseudonym certificates differ from standard X.509 certificates in that they omit or anonymize identifying fields, such as the Subject Name or Distinguished Name, to enhance user privacy and prevent direct correlation with a specific identity, while still enabling authentication and secure communication. Successful generation of certificates is backed by the software TPM of a legitimate client during initialization in a Nexagon. The TPM generates a fresh key pair and uses its private key to sign a request that includes the newly generated public key and an encrypted representation of its long-term identity. This request is sent to the CA, which verifies it with the TPM’s attestation. After validation, the CA issues a pseudonym certificate tied to the new key pair, containing metadata like permissions and expiration~\cite{9445398}.
\end{itemize}

\vspace{5pt}
Although other cryptographic methods such as group signatures~\cite{1181226} exist that guarantee privacy and accountability, pseudonym certificates offer better applicative advantages in Nexagons due to their scalability, flexibility, and enhanced privacy features. With pseudonym certificates individual-level anonymity is guaranteed through dynamic key rotation and omission of identifying fields, preventing re-identification and linkage attacks. 

\vspace{5pt}
\subsubsection{Mobile Client}
\begin{figure}[ht!]
    \centering
    \includegraphics[width=1.0\linewidth]{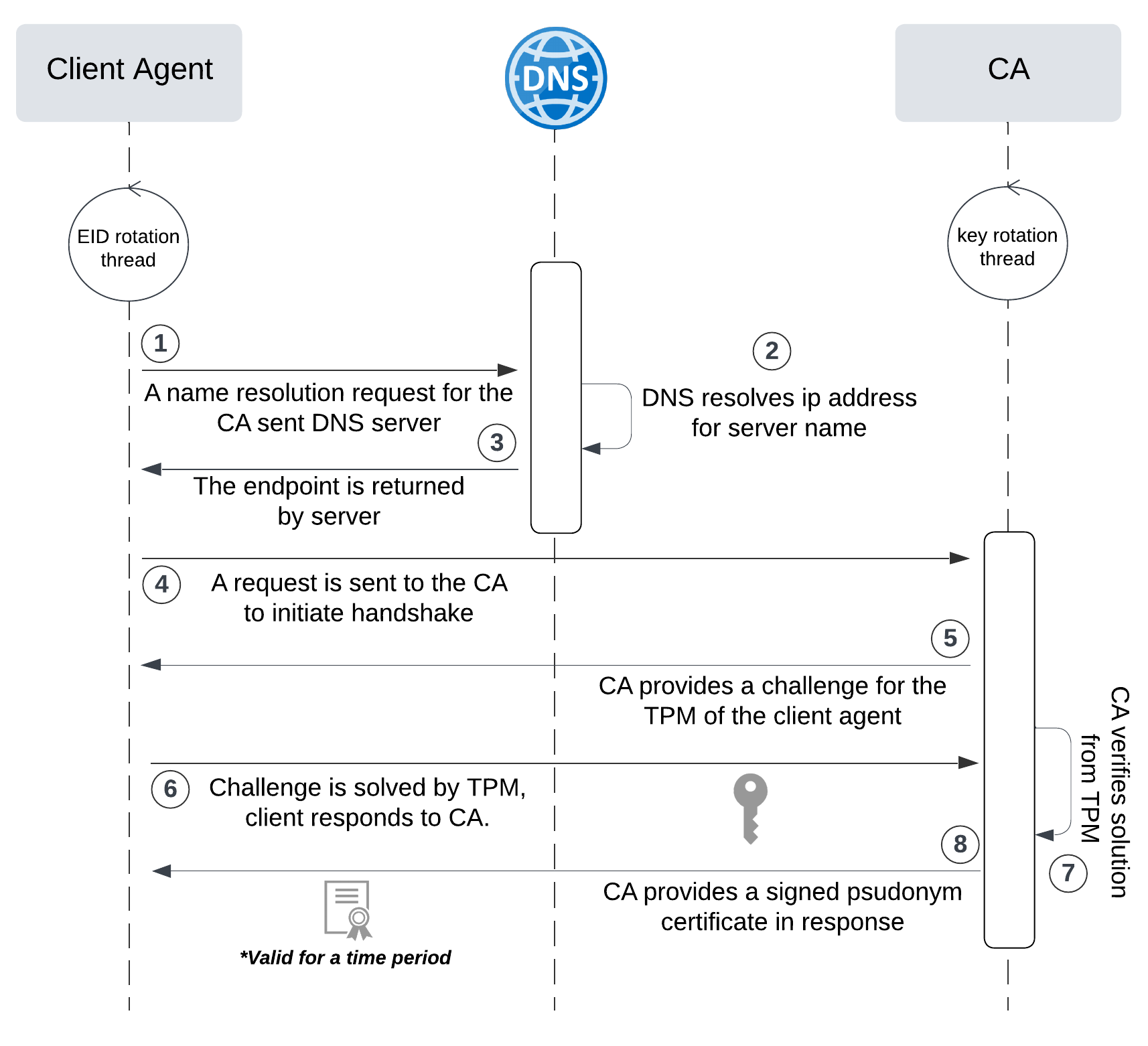}
    \vspace{-15pt}
    \caption{Sequence diagram illustrating the interaction between the client agent and the Certificate Authority (CA) during the initial onboarding phase.}
    \label{trust-ca}
    \vspace{-10pt}
\end{figure}

The mobile agent is capable of generating unique EIDs, which is essential to publish events while maintaining anonymity. After coming online, the client is configured to first resolve a pre-configured server name via DNS. It then contacts the authentication agent to obtain the necessary credentials.
Another significant addition is the ability for the agent to periodically swap EID when forwarding data across the network. The frequency of certificate swaps, combined with the randomness of these unique identifiers, effectively preserves the privacy of clients and their users. Figure~\ref{trust-ca} presents a sequence diagram that illustrates the steps undertaken by a mobile client during the initial onboarding process. This process is essential to ensure that the events published are acknowledged by the network.

\vspace{5pt}
\subsubsection{Mapping and Aggregation}
The mapping agent is equipped to store the current mappings of active clients as a key-value (KV) data store. We developed endpoints to enable clients to publish events when a detection occurs. The agent is also configured to forward events directly to the aggregation endpoints configured during authentication. The aggregation agent implements a streaming pipeline designed to process data transmitted by mapping agents when clients upload events.

\vspace{5pt}
\subsubsection{Environment}
We setup a test environment to assess the operation and performance of the Nexagon protocol. We used two virtual machines (VM), which were configured identically, each equipped with 4 GB of RAM and 2 CPU cores. This setup allowed us to run tests to mimic events from edge devices with and without the proposed security extensions. In the first VM, we ran a composite deployment of the entire management plane, with each component deployed as a microservice. The second VM hosted the client process, which was engineered to publish events that mimicked requests from mobile edge device. A comprehensive documentation of our architecture and codebase can be accessed in our repository~\cite{obadofin2024nexagon} upon request, providing further insight into our design and implementation choices.

\subsection{Performance Analysis}
In this section, our evaluations are presented together with the key results obtained during our mock tests.

\vspace{5pt}
\subsubsection{Evaluation Methodology}
We conducted load tests with varying numbers of user reqests made to compare the performance impact of our authentication extension. Initial tests established a baseline by processing client requests using a pre-shared key. Subsequent tests employed using our authentication mechanism as outlined previously. This approach enabled us to quantify the performance overhead introduced by the security enhancements. A summary of the key metrics captured are presented in Table~\ref{results}. In addition, Table~\ref{results2} provides a detailed breakdown of response times and their percentile distributions.

\vspace{5pt}
\subsubsection{Metrics}
The results indicate a latency increase of 10\% to 25\%. we also observed a slight decrease in throughput that ranged from 3\% to 7\%. These results are well within the acceptable limits for a real-world deployment.

\begin{table}[ht!]
\caption{Summary results for key metrics and confidence intervals}
\vspace{-5pt}
\begin{center}
\setlength{\tabcolsep}{6.5pt}
\renewcommand{\arraystretch}{1.5}
\begin{tabular}{|c|c|c|}
\hline
\textbf{} & \multicolumn{1}{|c|}{\textbf{Without Extension}} & \multicolumn{1}{|c|}{\textbf{With Extension}} \\
\hline
\textbf{Average Latency (ms)} & 306 (±73) & 384 (±45)  \\
\textbf{Throughput (req/sec)} & 260 (±19) & 250 (±10) \\
\textbf{ CPU Utilization (\%)} & 42 (±6) & 57 (±3) \\
\hline
\end{tabular}
\label{results}
\end{center}
\end{table}

\begin{table}[ht!]
\vspace{-5pt}
\caption{Distribution of latency at different percentiles}
\vspace{-5pt}
\begin{center}
\setlength{\tabcolsep}{6.5pt}
\renewcommand{\arraystretch}{1.5}
\begin{tabular}{|c|c|c|}
\hline
\textbf{}&\multicolumn{2}{|c|} 
{\textbf{Latency Percentiles (milliseconds)}} \\
\cline{2-3}
\textbf{} & \multicolumn{1}{|c|}{\textbf{Without Extension}} & \multicolumn{1}{|c|}{\textbf{With Extension}} \\
\hline
\textbf{50th Percentile} & 276 & 330 \\
\textbf{80th Percentile} & 290 & 373 \\
\textbf{90th Percentile} & 330 & 416 \\
\textbf{95th Percentile} & 400 & 460 \\
\hline
\end{tabular}
\label{results2}
\end{center}
\vspace{-5pt}
\end{table}

\section{conclusion}
Aiming at a safe, secure and efficient solution for crowdsourcing geo-referenced data in intelligent transportation systems, we delivered a detailed vulnerability analysis, mitigation strategies and a set of security extensions for IETF’s Nexagon protocol. Our prototype implementation takes advantage of concepts from the addressing capabilities of LISP and the indexing structure of H3. Our results show that a PKI with pseudo-random ephemeral certificates and identifiers delivers a robust solution that protects user privacy and the integrity of the Nexagon network without compromising latency and throughput. Future research could explore the deployment of the Nexagon protocol in real-world vehicular networks to evaluate its performance under dynamic conditions. In addition, advanced mechanisms such as federated learning for decentralized and privacy-preserving data processing can also be explored to further improve the feature set of the protocol and inform future studies.

\printbibliography

\end{document}